\documentclass[letterpaper,12pt]{article}
\usepackage[table]{xcolor}

\usepackage{amsmath}
\usepackage[ruled,vlined]{algorithm2e}
\usepackage{fancyvrb}
\usepackage{tikz-qtree}
\usepackage{tikz}
\usepackage{url}
\usetikzlibrary{trees}

\usepackage[margin=1in]{geometry}
\usepackage{mathptmx}
\usepackage[scaled=.90]{helvet}
\usepackage{courier}
\usepackage{setspace}
\usepackage{fancyhdr}
\usepackage{authblk}
\usepackage{amssymb}
\usepackage[table]{xcolor}
\usepackage{array}
\usepackage{mathtools}

\mathtoolsset{showonlyrefs}

\newcommand{\beginsupplement}{%
        \setcounter{table}{0}
        \renewcommand{\thetable}{S\arabic{table}}%
        \renewcommand{\thesection}{S}%
        \setcounter{figure}{0}
        \renewcommand{\thefigure}{S\arabic{figure}}%
     }

\tikzset{pics/tournament/.style={code={
   \def\pv##1{\pgfkeysvalueof{/tikz/tournament/##1}}    
   \tikzset{tournament/.cd,#1}  
   \foreach \XX [remember=\XX as \LastXX] in {\pv{n},...,0}
   {\pgfmathtruncatemacro{\mym}{pow(2,\XX)}
   \ifnum\XX=\pv{n}
    \foreach \YY in {1,...,\mym}
     {\draw (-\XX*\pv{d},{(\YY-\mym/2-1/2)*\pv{l}}) 
     -- ++ (\pv{d},0) |- coordinate[pos=0.25] (p-\XX-\YY)
     (-\XX*\pv{d},{(\YY-\mym/2-1)*\pv{l}})
     ;}
   \else
    \foreach \YY [evaluate=\YY as \ZZ using {int(2*\YY)}] in {1,...,\mym}
     {
     \draw (p-\LastXX-\ZZ) -- ++ (\pv{d},0)
     |- coordinate[pos=0.25] (p-\XX-\YY) 
     (p-\LastXX-\the\numexpr\ZZ-1);
     }
   \fi}
   \draw (p-0-1) -- ++ (\pv{d},0);
}},tournament/.cd,n/.initial=2,l/.initial=1,d/.initial=2}

\SetKwInput{Input}{Input}
\SetKwInput{Output}{Output}
\SetKwInput{Return}{Return}
\usepackage{xr}
\usepackage{xr-hyper}
\makeatletter
\newcommand*{\addFileDependency}[1]{
  \typeout{(#1)}
  \@addtofilelist{#1}
  \IfFileExists{#1}{}{\typeout{No file #1.}}
}
\makeatother

\title{Using Conformal Win Probability to Predict the Winners of the Cancelled 2020 NCAA Basketball Tournaments}
\author[1]{Chancellor Johnstone}
\author[2]{Dan Nettleton}
\affil[1]{Air Force Institute of Technology, Department of Mathematics and Statistics}
\affil[2]{Iowa State University, Department of Statistics}
\date{\today}                     
\renewcommand\Affilfont{\itshape\small}

\SetKwInput{Input}{ Input}
\SetKwInput{Output}{Output}
\SetKwInput{Return}{Return}

\numberwithin{theorem}{section}
\numberwithin{definition}{section}

\usepackage{mathptmx}

\usepackage{graphicx}

\usepackage[authoryear,comma,longnamesfirst,sectionbib]{natbib} 

\usepackage[hidelinks]{hyperref}

\begin{document}
\setcitestyle{aysep={},yysep={;}}
\maketitle
\thispagestyle{empty}

\begin{abstract}
The COVID-19 pandemic was responsible for the cancellation of both the men's and women's 2020 National Collegiate Athletic Association (NCAA) Division 1 basketball tournaments. Starting from the point at which the Division 1 tournaments and any unfinished conference tournaments were cancelled, we deliver closed-form probabilities for each team of making the Division 1 tournaments, had they not been cancelled, aided by use of conformal predictive distributions. We also deliver probabilities of a team winning March Madness, given a tournament bracket. We then compare single-game win probabilities generated with conformal predictive distributions, aptly named conformal win probabilities, to those generated through linear and logistic regression on seven years of historical college basketball data, specifically from the 2014-2015 season through the 2020-2021 season. Conformal win probabilities are shown to be better calibrated than other methods, resulting in more accurate win probability estimates, while requiring fewer distributional assumptions. 
\newline
\newline
\noindent
Keywords: Conformal inference, predictive distributions, sports analytics, uncertainty quantification.
\end{abstract}

\clearpage
\pagenumbering{arabic}

\section{Introduction}
\label{sec:intro}
Two of the most popular tournaments in the world are the men's and women's National Collegiate Athletic Association (NCAA) Division 1 basketball tournaments. In college basketball, teams are grouped into conferences. Over the course of the regular season, teams compete against opponents within their own conference as well as teams outside their conference. Following the regular season, better performing teams within each conference compete in a conference tournament, with the winner of said tournament earning an invitation to play in the Division 1 tournament. The invitation for winning a conference tournament is called an ``automatic bid". Historically, sixty-four teams are selected for the women's tournament. Thirty-two of the sixty-four teams are automatic bids, corresponding to the thirty-two conference tournament winners. The other thirty-two teams are ``at-large bids", made up of teams failing to win their respective conference tournament. At-large bids are decided by a selection committee, which has guidelines that govern how to choose not only the teams invited to the tournament, but also how to set the tournament bracket, which defines who and where each team will play initially and could play eventually. Teams that earn an automatic bid or an at-large bid are said to have ``made the tournament".

In previous iterations of March Madness, the men's tournament has differed slightly from the women's tournament, with the former including a set of games called the First Four. In the First Four, eight teams compete for four spots in the round of 64, also called the First Round. Specifically, the four lowest ranked automatic bids compete for two spots in the First Round, while the four lowest ranking at-large bids compete against each other for the two remaining spots. Thus, the men's tournament includes thirty-two automatic bids and thirty-six at-large bids. In 2022, the women's tournament included a First Four for the first time in tournament history, bringing the total number of teams in the tournament up to sixty-eight \citep{nytfirstfour}. Another difference between the men's and women's tournaments is that the NCAA has historically referred to the men's (but not the women's) tournament as ``March Madness" \citep{ncaabrand}. In this paper, we use the term to describe both the men's and women's tournament.

As a result of the COVID-19 pandemic, the NCAA cancelled both the men's and women's 2020 NCAA tournaments. A majority of athletic conferences followed by cancelling their own conference tournaments, leaving many automatic bids for March Madness undecided. Due to these cancellations, natural questions arise with respect to which teams might have made the March Madness field and which teams might have won the tournament, if it had occurred. Using data from the 2019-2020 men's and women's collegiate seasons, we deliver probabalistic answers to these questions. 

Specifically, we contribute the following: 1) an overall ranking of Division 1 teams, as well as estimates of each team's strength, based on 2019-2020 regular season data, 2) closed-form calculations for probabilities of teams making the 2019-2020 March Madness field, calculated beginning from the point in time at which each conference tournament was cancelled, under a simplified tournament selection process, 3) closed-form calculations of probabilities of teams winning March Madness, given each of several potential brackets, and 4) a new pair of fully audited data sets with observed margins of victory for both men's and women's Division 1 basketball, spanning from the 2014-2015 season through the 2020-2021 season.

The calculation of probabilities for teams making the 2019-2020 March Madness field consider each conference tournament's unfinished bracket as well as our estimates of Division 1 team strengths, which we fix following the culmination of the regular season. The closed-form nature of the probabilities also reduces the computational load and eliminates error inherent to simulation-based approaches. To our knowledge, this is the first closed-form approach to take into account partially completed conference tournaments when generating probabilities of making the March Madness field.

Estimating March Madness win probabilities prior to the selection of the tournament field and the determination of the March Madness bracket is a difficult problem. If we define all the potential brackets as the set $\mathcal{B}$, we can decompose the probability of a team winning March Madness as

\begin{equation}
\label{eqn:b-wp}
    \mathbb{P}(W_u = 1) = \sum_{B \in \mathcal{B}} \mathbb{P}(W_u = 1|B)\mathbb{P}(B),
\end{equation}

\noindent
where $\{W_u = 1\}$ represents team $u$ winning March Madness. However, calculations for all possible brackets are intractable. For a set of, say, 350 teams, there are $\binom{350}{64}$ ways to select a field of teams to compete in a 64-team tournament. Given a tournament field of $N = 2^J$ teams, where $J$ is the number of rounds in the tournament ($J = 6$ for a 64-team tournament), the number of unique brackets for a single-elimination tournament is

\begin{equation}
    \prod_{i = 1}^{N/2} \binom{2i}{2}\bigg/2^{N/2-1}, 
    \label{eqn:num-bracks}
\end{equation}


\noindent
which grows rapidly as $N$ increases. An 8-team tournament results in 315 potential brackets, while a $16$-team tournament results in 638,512,875 potential brackets. In the case of March Madness, the size of the set $\mathcal{B}$ is enormous.

Of course, some brackets are more likely than others due to the set of constraints used by the selection committee. However, even if the set of plausible brackets for March Madness was small relative to the complete set $\mathcal{B}$ when the tournaments were cancelled in 2020, estimating $\mathbb{P}(B)$ in \eqref{eqn:b-wp} for any given bracket $B$ depends on the complex and, ultimately, subjective decision making process used by the NCAA selection committee. Thus, we make no attempt to estimate $\mathbb{P}(B)$ for any bracket $B$. Instead, in this paper, we focus on the construction of the marginal probability of each team making the March Madness field. Additionally, using brackets suggested by experts, along with brackets we construct, we compare March Madness win probabilities, $\mathbb{P}(W_u = 1|B)$ for all teams $u$, across different brackets $B$. We find that the win probabilities for teams most likely to win are relatively stable across brackets. Baylor, South Carolina, and Oregon each had more than a 20\% win probability for most of the brackets we considered for the women's tournament. On the men's side, Kansas was the most likely to win the tournament regardless of the bracket.

Another contribution of the paper is the novel application of conformal predictive distributions \citep{vovk2019nonparametric} for the estimation of win probabilities. Conformal predictive distributions allow for the construction of win probability estimates under very mild distributional assumptions, reducing dependence on normality assumptions for our results. We find that conformal predictive distributions provide win probability estimates that are superior to other methods relying on stronger assumptions when compared using seven years of men's and women's post-season NCAA basketball data. 

Section \ref{sec:dfpi-pred-sports} provides background on constructing overall win probabilities for single-elimination tournaments and introduces the closed-form calculation of probabilities related to March Madness. Section \ref{sec:wp-methods} describes three methods for generating win probabilities of individual games, including the construction of win probability estimates through conformal predictive distributions. Section \ref{sec:results} describes the overall results, to include a ranking of the top teams, conference tournament and March Madness win probabilities associated with the 2019-2020 NCAA Division 1 basketball season and a comparison of three win probability generation methods. Section \ref{sec:conclusion} concludes the paper. All of the R code and data sets used in this research are available at

\begin{center}
    \url{https://github.com/chancejohnstone/marchmadnessconformal}. 
\end{center}

\section{Probabilities for March Madness}
\label{sec:dfpi-pred-sports}

In the following section, we describe win probability as it relates to single-elimination tournaments like March Madness. We also introduce the probability of a team making the March Madness field, given a collection of conference tournament brackets, team rankings and game-by-game win probabilities. We limit our discussion scope in this section primarily to the women's tournament, but the general construction reflects the men's tournament as well.

Throughout this paper, we use the common verbiage that a team is ranked ``higher" than another team if the former team is believed to be better than the latter team. Likewise, a ``lower" ranking implies a weaker team. We follow the common convention that a team of rank $r$ has a higher rank than a team of rank $r+s$ for $s>0$. Teams ranked 1 to 32 are collectively identified as ``high-ranked". Teams ranked below 64 are identified as ``low-ranked". While the colloquial use of the term ``bubble teams" is usually reserved to describe a subset of teams near the boundary separating teams in and out of the March Madness field, we use the term to explicitly describe the teams ranked 33 to 64. In Section \ref{sec:sports-app}, we discuss an approach to rank teams based on observed game outcomes.

\subsection{Win Probability for Single-Elimination Tournaments}

Suppose were are given a game between team $u$ and team $v$ with the win probability for team $u$ defined as $p_{uv}$. While the true value of $p_{uv}$ is not known in practice, we describe methods for estimating probabilities for any match-up in Section \ref{sec:wp-methods}. Given these probabilities, one method for providing estimates of overall tournament win probability is through simulation. We can simulate the outcome of a game between team $u$ and team $v$ by randomly sampling from a standard uniform distribution. A value less than $p_{uv}$ corresponds to a victory for team $u$, while a value greater than $p_{uv}$ represents a victory for team $v$. Every game in a tournament can be simulated until we have an overall winner. We can then repeat the entire simulation process multiple times to get a Monte Carlo estimate of each team's probability of winning said tournament. 

While a simulation-based approach is effective at providing estimates of the true tournament win probability for each team, simulation requires excessive computational effort, with each estimate having inherent Monte Carlo error. To eliminate Monte Carlo error, we can generate overall tournament win probabilities through closed-form calculation. 

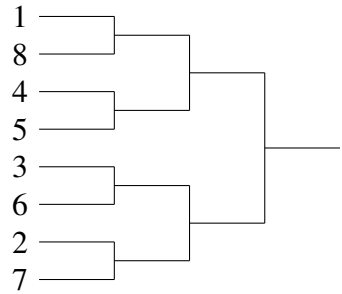
\begin{figure}
\centering
\begin{tikzpicture}
    \draw  (0,0) -- (-1,0) node[left]{1};
    \draw  (0,-.5) -- (-1,-.5) node[left]{8};
    \draw  (0,-1) -- (-1,-1) node[left]{4};
    \draw  (0,-1.5) -- (-1,-1.5) node[left]{5};
    \draw  (0,-2) -- (-1,-2) node[left]{3};
    \draw  (0,-2.5) -- (-1,-2.5) node[left]{6};
    \draw  (0,-3) -- (-1,-3) node[left]{2};
    \draw  (0,-3.5) -- (-1,-3.5) node[left]{7};
    
    \draw  (0,0) -- (0,-.5);
    \draw  (0,-1) -- (0,-1.5);
    \draw  (0,-2) -- (0,-2.5);
    \draw  (0,-3) -- (0,-3.5);
    
    \draw  (1,-.25) -- (0,-.25);
    \draw  (1,-1.25) -- (0,-1.25);
    \draw  (1,-2.25) -- (0,-2.25);
    \draw  (1,-3.25) -- (0,-3.25);
    
    \draw  (1,-.25) -- (1,-1.25);
    \draw  (1,-2.25) -- (1,-3.25);
    
    \draw  (2,-.75) -- (1,-.75);
    \draw  (2,-2.75) -- (1,-2.75);
    
    \draw  (2,-.75) -- (2,-2.75);
    
    \draw  (3,-1.75) -- (2,-1.75);
\end{tikzpicture}
\caption{Bracket for eight-team single-elimination tournament.}
\label{fig:8bracket}
\end{figure}

Suppose we have an eight team single-elimination tournament with the bracket shown in Figure \ref{fig:8bracket}. The highest ranking team, team 1, plays the lowest ranking team, team 8, in the first round. Assuming team 1 was victorious in round one, their second round opponent could be team 4 or 5. In the third round, team 1 could play team 3, 6, 2 or 7. After the first round of the tournament, team 8 has the same potential opponents as team 1.

Using the knowledge of a team's potential opponents in future games, we can calculate win probabilities for any upcoming round and, thus, the entire tournament. Formalized in \cite{edwards1991combinatorial}, the tournament win probability for team $u$ given a fixed, single-elimination tournament bracket with $J$ rounds is

\begin{equation}
    \label{eqn:cf-wp}
    q_{uJ} = q_{uJ-1}\Bigg[ \sum_{s \in \mathcal{O}_{uJ}} p_{us} q_{sJ-1} \Bigg],
\end{equation}

\noindent
where $q_{uj}$ is the probability that team $u$ wins in round $j = 1,\hdots,J$, and $\mathcal{O}_{uj}$ is the set of potential opponents team $u$ could play in round $j$. We explicitly set $q_{u1} = p_{u\mathcal{O}_{u1}}$, where $\mathcal{O}_{u1}$ is team $u$'s opponent in round one. We can extend \eqref{eqn:cf-wp} to single-elimination tournaments of any size or construction as long we are able to determine the set $\mathcal{O}_{uj}$ for any team $u$ in any round $j$. 

\subsection{Probability for Making the NCAA Tournament}
\label{sec:making} 

With \eqref{eqn:cf-wp} we can generate an overall tournament win probability for each team in a tournament exactly, given a fixed tournament bracket and game-by-game win probabilities. However, following the regular season, but prior to the culmination of all conference tournaments, the field for March Madness is not fully known. Thus, we cannot utilize \eqref{eqn:cf-wp} directly for estimating team win probabilities for the 2020 March Madness tournament. We first turn our attention to estimating each women's team's probability of making the 2020 March Madness field, made up of thirty-two automatic bids and thirty-two at-large bids. Although the closed-form calculations reflect probabilities related to the 2019-2020 women's March Madness tournament, which did not include a First Four, only slight changes are required to reflect the inclusion of a First Four for the men's and future women's tournaments. 

We define $F_u$ as the indicator variable for whether or not the $u$-th ranked team makes the NCAA tournament field. Knowing that the NCAA tournament is made up of automatic and at-large bids, we define two relevant indicator variables $C_u$ and $L_u$ associated with a team receiving one of these bids, respectively. $C_u$ is one if team $u$ wins its conference tournament and zero otherwise. We define $L_u$ as the number of conference tournaments won by teams ranked below team $u$. Then, under the assumption that higher-ranked at-large bids make the March Madness field before lower-ranked at-large bids, for any team $u$, the probability of making the NCAA tournament is

\begin{equation}
\mathbb{P}(F_u = 1) = \mathbb{P}(\{C_u = 1\} \cup \{L_u \le t_u\}) = \mathbb{P}(C_u = 1) + \mathbb{P}(L_u \le t_u) - \mathbb{P}(C_u = 1, L_u \le t_u),
\label{eqn:out-64}
\end{equation}

\noindent
where $t_u = 64 - u$ is the maximum number of teams ranked below team $u$ that can receive an automatic bid without preventing team $u$ from making the NCAA tournament as an at-large bid. Because there are only 32 conference tournaments, $L_u$ is less than or equal to 32 with probability one. Thus, with the current construction, teams ranked 32 or higher always make the NCAA tournament. For low-ranked teams, \eqref{eqn:out-64} reduces to $\mathbb{P}(C_u = 1)$, aligning with the fact that weaker teams must win their conference tournament to get an invite to March Madness. 

We can decompose the intersection probability of \eqref{eqn:out-64} into

\begin{equation}
\label{eqn:cond-T}
\mathbb{P}(C_u = 1, L_u \le t_u) = \mathbb{P}(L_u \le t_u|C_u = 1)\mathbb{P}(C_u = 1).
\end{equation}

\noindent
To explicitly describe the probabilities in \eqref{eqn:cond-T}, we split the teams in each conference into two sets, $\mathcal{H}^u_k$ and $\mathcal{L}^u_k$, defining $\mathcal{H}^u_k$ as the set of teams in conference $k = 1, \hdots, K$ ranked higher than or equal to team $u$ and $\mathcal{L}^u_k$ as the set of teams in conference $k$ ranked lower than team $u$. We reference lower or higher-ranked teams in the same conference as team $u$ using $k(u)$ instead of $k$. It is important to emphasize that team $u$ is included in $\mathcal{H}^u_{k(u)}$. Let $C_{\mathcal{H}_k^u}$ be one if a team in $\mathcal{H}_k^u$ wins conference tournament $k$ and zero otherwise. $C_{\mathcal{L}_k^u}$ is defined in a similar manner. 

We assume that the outcome of any conference tournament is independent of the outcome of any other conference tournament. Thus, we can describe $L_u$ as a sum of independent, but not identically distributed, Bernoulli random variables,

\begin{equation}
\label{eqn:out-64-sum-k}
L_u = \sum_{k = 1}^K C_{\mathcal{L}^u_k}.
\end{equation}

\noindent
If $C_{\mathcal{L}^u_k}$ were identically distributed for all conferences, then $L_u$ would be a binomial random variable. Because this not the case, $L_u$ is instead a Poisson-binomial random variable with cumulative distribution function

\begin{equation}
\label{eqn:poisson-binom}
\mathbb{P}(L_u \le l) = \sum_{m = 0}^{l} \Bigg\{ \sum_{A \in \mathcal{F}_m} \prod_{s \in A} p_s \prod_{s \in A^C} (1 - p_s)\Bigg\},
\end{equation}

\noindent
where $p_k$ is the probability of a team in $\mathcal{L}^u_k$ winning conference tournament $k$, and $\mathcal{F}_m$ is the set of all unique $m$-tuples of $\{1,\hdots,32\}$. With \eqref{eqn:poisson-binom} known, the conditional portion of \eqref{eqn:cond-T} is a new Poisson-binomial random variable where $p_{k(u)} = 0$ because we condition on team $u$ winning their conference tournament. Thus, the probability of team $u$ making the tournament is

\begin{equation}
\label{eqn:t1}
\mathbb{P}(F_u = 1) = q_{uJ_{k(u)}} + \mathbb{P}(L_u \le t_u) - \Bigg(\sum_{m = 0}^{t_u} \Bigg\{ \sum_{A \in \mathcal{F}_m} \prod_{s \in A} p'_s \prod_{s \in A^C} (1 - p'_s)\Bigg\} \Bigg) \times q_{uJ_{k(u)}},
\end{equation}

\noindent
where $p'_{k}$ is equal to $p_k$ when $k$ is not equal to $k(u)$ and zero otherwise, and $J_{k(u)}$ is the number of rounds in the conference tournament for conference $k(u)$.

While the above derivation provides a closed-form calculation for probabilities of making the March Madness field, it does not describe any team's probability of winning March Madness. To do this, we must also derive closed-form probability calculations for specific tournament brackets. However, as discussed in Section \ref{sec:intro}, it is difficult to explicitly construct calculations for this task due to the inherent subjectivity associated with the seeding of teams. For this reason, we include the derivation of the closed-form marginal probability calculation for a team's March Madness rank under an adjusted tournament selection process utilizing the S-curve method \citep{ncaa2021} in Supplementary Materials.

\section{Win Probabilities for Individual Games}
\label{sec:wp-methods}

Determining win probability in sports primarily began with baseball \citep{lindsey1961progress}. Since then, win probability has permeated many sports and become a staple for discussion among sports analysts and enthusiasts. Example applications of win probability have been seen in sports such as basketball \citep{stern1994brownian, loeffelholz2009nba}, hockey \citep{gramacy9estimating}, soccer \citep{hill1974association, karlis2008bayesian, robberechts2019will}, football \citep{stern1991probability, lock2014nflwp}, cycling \citep{moffatt2014lead}, darts \citep{liebscher2017predicting}, rugby \citep{lee1999applications}, cricket \citep{asif2016play}, table tennis \citep{liu2016new} and even video games \citep{semenov2016performance}. 

A majority of these methodologies use some form of parametric regression to capture individual and/or team strengths, offensive and/or defensive capabilities or other related effects. We continue the parametric focus by using a linear model framework to estimate team strengths, but our proposed approach makes only minimal closed-form distributional assumptions. 

Initially, suppose that

\begin{equation}
y_i = x_i'\beta + \epsilon_i,
\label{eqn:lin-model}
\end{equation}

\noindent
where $y_i$ represents the response of interest for observation $i$, $x_i$ is a length $p$ vector of covariates for observation $i$, $\beta$ is the vector of true parameter values and $\epsilon_i$ is a mean-zero error term. We define $y = (y_1, \hdots, y_n)'$ and $X = (x_1, \hdots, x_n)'$, where the vector $y$ and matrix $X$ make up our $n$ observations $D_n = \{(x_i, y_i)\}_{i=1}^n$. We are interested in both predicting $y_{n+1}$ and quantifying uncertainty about $y_{n+1}$, given $x_{n+1}$, for some new observation $(x_{n+1}, y_{n+1})$. In subsequent sections, the response values in $y$ will be margins of victory, and the elements of $\beta$ will include team strength parameters. However, at this stage a slightly more general treatment is useful.

In the following section, we discuss event probability estimation using three different methods: conformal predictive distributions based on model \eqref{eqn:lin-model}, linear regression with model \eqref{eqn:lin-model} and an added assumption of mean-zero, normally distributed, independent errors, and logistic regression. We then provide specific application to the sports context, extending the aforementioned methods in order to estimate win probabilities in sports.

\subsection{Event Probability with Conformal Predictive Distributions}
\label{sec:conf-event}

Predictive distributions, e.g., those introduced in \cite{lawless2005frequentist}, provide a method for estimating the conditional distribution of a future observation given observed data. Conformal predictive distributions (CPDs) \citep{vovk2019nonparametric} provide similar results but through the use of a distribution-free approach based on conformal inference \citep{gammerman1998learning}. In the following section we provide a general treatment of conformal inference, followed by an introduction to conformal predictive distributions.

\subsubsection{Conformal Inference}

The aim of conformal inference is to quantify uncertainty in classification and/or regression tasks under weak distributional assumptions. In a regression context, conformal inference produces conservative prediction intervals for some unobserved response $y_{n+1}$ through the repeated inversion of some hypothesis test, say

\begin{equation}
H_0: y_{n+1} = y_c  \; \textrm{ vs. } \; H_a: y_{n+1} \ne y_c,
\label{eqn:conf-permute-2}
\end{equation}

\noindent
where $y_{n+1}$ is the response value associated with an incoming covariate vector $x_{n+1}$, and $y_c$ is a candidate response value \citep{lei2018distribution}. The only assumption required to achieve valid prediction intervals is that the data $D_n$ combined with the new observation $(x_{n+1}, y_{n+1})$ comprise an exchangeable set of observations. 

The inversion of \eqref{eqn:conf-permute-2} is achieved through refitting the model of interest with an augmented data set that includes the data pair $(x_{n+1}, y_c)$. For each candidate value, a set of \textit{conformity scores} is generated, one for each observation in the augmented data set. A conformity score measures how well a particular data point conforms to the rest of the data set and traditionally utilizes the data pair $(x_i,y_i)$ and the prediction for $y_i$, denoted $\hat{y}_i(y_c)$, as arguments. While the prediction $\hat{y}_i(y_c)$ is dependent on both $(x_{n+1}, y_c)$ and $D_n$, we omit dependence on $x_{n+1}$ and $D_n$ in our notation. We define

\begin{equation}
\pi(y_c, \tau) = \frac{1}{n+1} \sum_{i = 1}^{n+1} \Big[\mathbb{I}\{R_i(y_c) < R_{n+1}(y_c)\} + \tau\mathbb{I}\{R_i(y_c) = R_{n+1}(y_c)\}\Big],
\label{eqn:conf-p-values-2}
\end{equation}

\noindent
where, for $i = 1, \hdots, n$, $R_i(y_c)$ is the conformity score for the data pair $(x_i, y_i)$ as a function of $(x_{n+1}, y_c)$, $R_{n+1}(y_c)$ is the conformity score associated with $(x_{n+1},y_c)$, and $\tau$ is a $U(0,1)$ random variable. 

In hypothesis testing we generate a probability associated with an observed test statistic, specifically the probability of a \textit{more extreme} value than the observed test statistic under the assumption of a specified null hypothesis, also known as a $p$-value. With the construction of $\pi(y_c, \tau)$, we generate an estimate of the probability of an observation \textit{less extreme} than the candidate value $y_c$. Thus, $1 - \pi(y_c, \tau) $ provides a $p$-value associated with \eqref{eqn:conf-permute-2} \citep{shafer2008tutorial, lei2018distribution}. The inclusion of the random variable $\tau$ generates a smoothed conformal predictor \citep{vovk2005algorithmic}. 

For a fixed $\tau$, we can construct a conformal prediction region for the response associated with $x_{n+1}$,

\begin{equation}
C_{1-\alpha, \tau}(x_{n+1}) = \{y_c \in \mathbb{R} \; : \; (n+1) \pi(y_c, \tau) \le \lceil(1 - \alpha)(n+1) \rceil \},
\label{eqn:conf-pi-2}
\end{equation}

\noindent
where $1-\alpha$ is the nominal coverage level. When $\tau$ is one, $\pi(y_c, 1)$ is the proportion of observations in the augmented data set whose conformity score is less than or equal to the conformity score associated with candidate value $y_c$. Regardless of the conformity score, a conformal prediction region with nominal coverage level $1-\alpha$ is conservative \citep{vovk2005algorithmic}. Thus, for some new observation $(x_{n+1},y_{n+1})$,

\begin{equation}
    \label{eqn:conf-prob}
    \mathbb{P}\big(y_{n+1} \in  C_{1-\alpha, \tau}(x_{n+1})\big) \ge 1 - \alpha.
\end{equation}

\noindent

\subsubsection{Conformal Predictive Distributions}
\label{sec:cpds}

In the previous section we explained conformal inference in general terms. However, we can construct $\pi(y_c, \tau)$ with certain conformity scores to achieve inference for different events associated with $y_{n+1}$. One commonly used conformity score in a regression setting is the absolute residual, $|y_i - \hat{y}_i(y_c)|$, which leads to symmetric prediction intervals for $y_{n+1}$ around a value $\tilde{y}$ satisfying $\tilde{y} = \hat{y}_{n+1}(\tilde{y})$. The traditional residual associated with a prediction, $y_i - \hat{y}_i(y_c)$, results in a one-sided prediction interval for $y_{n+1}$ of the form $\big(-\infty, u(D_n, x_{n+1}) \big)$. Additionally, the selection of the traditional residual as our conformity score turns $\pi(y_c, \tau)$ into a conformal predictive distribution \citep{vovk2019nonparametric}, which provides more information with respect to the behavior of random variables than, say, prediction intervals. For example, with a CPD, we can provide an estimate of the probability of the event $y_{n+1} \le y^*$. For the the remainder of this paper we construct $\pi(\cdot, \tau)$ using the conformity score $R_i(y_c) = y_i - \hat{y}_i(y_c)$.

As previously stated, $1-\pi(y_c, \tau)$ provides a $p$-value associated with \eqref{eqn:conf-permute-2}. Thus, $1-\pi(y_c,1/2)$ is analogous to the mid $p$-value, which acts a continuity correction for tests involving discrete test statistics. We point the interested reader to \cite{lancaster1949combination, lancaster1961significance}, \cite{ barnard1989alleged} and \cite{routledge1992resolving} for additional details on the mid $p$-value. We set $\tau = 1/2$ for the computation of our conformal predictive distributions throughout the remainder of this paper.

While we have generalized conformal predictive probabilities for the event $y_{n+1} \le y^*$, we focus on the case where $y^*$ is equal to zero in later sections and instead describe probabilities associated with the event $y_{n+1} > 0$, which represent win probabilities when $y_{n+1}$ is a margin of victory.

\subsection{Other Event Probability Methods}
\label{sec:other-wp-methods}

We specifically outline two competing methods to conformal predictive distributions: event probability through linear regression and event probability through logistic regression.

\subsubsection{Event Probability Through Linear Regression}
\label{sec:norm-wp}

We can estimate the expected value of some new observation $y_{n+1}$ using \eqref{eqn:lin-model}, but additional assumptions are required to provide event probabilities. In linear regression, the error term $\epsilon_{i}$ is traditionally assumed to be a mean-zero, normally distributed random variable with variance $\sigma^2 < \infty$. Together, these assumptions with independence among error terms make up a Gauss-Markov model with normal errors (GMMNE). 

A least-squares estimate for the expectation of $y_{n+1}$, $\hat{y}_{n+1}$, is $x'_{n+1}\hat{\beta}$ where $\hat{\beta} = (X'X)^{-1}X'y$ when $X$ is a full rank $n \times p$ matrix of covariates. Given the assumption of a GMMNE, $\hat{y}_{n+1}$ is normally distributed with mean $x'_{n+1}\beta$ and variance $\sigma^2(x'_{n+1}(X'X)^{-1}x_{n+1})$. The prediction error for observation $n+1$, $r_{n+1} = y_{n+1} - \hat{y}_{n+1}$, is also normally distributed with mean zero and variance $\sigma^2(1 + x'_{n+1}(X'X)^{-1}x_{n+1})$. Dividing $r_{n+1}$ by its estimated standard error then yields a $t$-distributed random variable. Thus, we can describe probabilities for events of the form $y_{n+1} > s$ using the standard predictive distribution

\begin{equation}
    \mathbb{P}(y_{n+1} > s) = 1 - F_{t,n-p}\Bigg(\frac{s-\hat{y}_{n+1}}{\hat{\sigma}\sqrt{1+ x'_{n+1}(X'X)^{-}x_{n+1}}}\Bigg), 
    \label{eqn:pivot}
\end{equation}

\noindent
where $\hat{\sigma}^2 = y'(I - X'(X'X)^{-1}X'y/(n-p)$ is the usual unbiased estimator of the error variance $\sigma^2$, and $F_{t,n-p}$ is the cumulative distribution function for a $t$-distributed random variable with $n - p$ degrees of freedom \citep{wang2012fiducial, vovk2019nonparametric}.

\subsubsection{Event Probability Through Logistic Regression}

While linear regression allows for an estimate of $\mathbb{P}(y_{n+1} > 0)$ based on assumptions related to the random error distribution, we can also generate probability estimates explicitly through logistic regression. Suppose we still have observations $D_n$. We define a new random variable $z_i$ such that $z_i = \mathbb{I}\{y_i > 0\}$. Instead of assumptions related to the distribution of the random error term $\epsilon_i$, we assume a relationship between the expectation of $z_i$, defined as $p_i$, and the covariates $x_i$ such that $\log \big( \frac{p_i}{1-p_i} \big) = x_i'\beta$. Then, we can then derive an estimate for $p_i$,

\begin{equation}
\label{eqn:bradley-probs}
    \hat{p}_{i} = \frac{e^{x_{i}'\hat{\beta}}}{1 + e^{x_{i}'\hat{\beta}}},
\end{equation}

\noindent
where $\hat{\beta}$ is the maximum-likelihood estimate for $\beta$ under the assumption that $z_1, \hdots, z_n$ are independent Bernoulli random variables. 

\subsection{Application to Win Probability in Sports}
\label{sec:sports-app}

We now extend the methods outlined in Section \ref{sec:conf-event} and Section \ref{sec:other-wp-methods} to a sports setting for the purpose of generating win probabilities. Specifically, we wish to identify win probabilities for some future game between a home team $u$ and away team $v$. 

The method of generating win probabilities in our case are made possible through the estimation of team strengths. One of the earliest methods for estimating relative team strength comes from \cite{harville1977ranks, harville1980predictions}, which uses the \textit{margin of victory} (MOV) for each game played. We focus on the initial linear model

\begin{equation}
y_{uv} = \mu + \theta_u - \theta_v + \epsilon_{uv},
\label{eqn:lin-model-sports1}
\end{equation}

\noindent
where $y_{uv}$ represents the observed MOV in a game between team $u$ and $v$ ($u \ne v$), with the the first team at home and the second away, $\theta_u$ represents the relative strength of team $u$ across a season, $\mu$ can be interpreted as a ``home court" advantage parameter, and $\epsilon_{uv}$ is a mean-zero error term. We can align \eqref{eqn:lin-model-sports1} with \eqref{eqn:lin-model} and identify games across different periods, e.g., games happening in a given week, by assuming

\begin{equation}
y_{uvw} = x_{uvw}'\beta + \epsilon_{uvw},
\label{eqn:lin-model-sports}
\end{equation}

\noindent
where $y_{uvw}$ represents the observed MOV in a game between team $u$ and $v$ ($u \ne v$) in period $w$, $\beta$ is the parameter vector $(\mu, \theta_1,\hdots,\theta_{p-1})'$, $\epsilon_{uvw}$ is a mean-zero error term, and $x_{uvw}$ is defined as follows. For $i = 1, \hdots, p$, let $e_t$ be the $t$-th  column of the $p \times p$ identity matrix, and let $e_{p+1}$ be the $p$-dimensional zero vector. Then, $x_{uvw}= e_1 + e_{u+1} - e_{v+1}$ for a game played on team $u$'s home court or $x_{uvw}= e_{u+1} - e_{v+1}$ for a game played at a neutral site.

Without loss of generality, we estimate team strengths under model \eqref{eqn:lin-model-sports} relative to an arbitrarily chosen baseline team. Let $\hat{\theta}_u$ be element $u$ + 1 of the least squares estimate for $\beta$ under model \eqref{eqn:lin-model-sports}, and define $\hat{\theta}_p = 0$. Then $\hat{\theta}_u - \hat{\theta}_v$ is the estimated margin of victory for team $u$ in a neutral-site game against team $v$, and $\hat{\theta}_1, \hdots, \hat{\theta}_p$ serve as estimated strengths of teams $1, \hdots, p$, respectively. The rank order of these estimated team strengths provides a ranking of the $p$ teams. 

By the definition of $y_{uvw}$, the probability that $y_{uvw}$ is greater than zero is the probability of a positive MOV, representing a win for the home team. Thus, with the assumption of \eqref{eqn:lin-model-sports}, we can now describe the event probability methods outlined in Section \ref{sec:conf-event} and Section \ref{sec:other-wp-methods} as they relate to win (and loss) probabilities in sports.

The different model assumptions do not change the inherent construction of event probability estimates with CPDs. We can align CPDs with model \eqref{eqn:lin-model-sports} by defining


\begin{equation}
\pi_w(y_c, \tau) = \frac{1}{n_w+1} \sum_{(u,v,w)} \Big[\mathbb{I}\{R_{uvw}(y_c) < R_{n_w+1}(y_c)\} + \tau\mathbb{I}\{R_{uvw}(y_c) = R_{n_w+1}(y_c)\}\Big],
\label{eqn:cpd-week}
\end{equation}

\noindent
where $n_w$ is the number of observations up to and including period $w$, $x_{n_w + 1}$ is the covariate vector associated with our game of interest, $R_{uvw}(y_c)$ is constructed using the using the prediction $\hat{y}_{uvw}(y_c)$ and $R_{n_w + 1}(y_c)$ is the conformity score associated with $(x_{n_w + 1}, y_c)$. We call the construction of win probability through CPDs \textit{conformal win probability}. As discussed in Section \ref{sec:cpds}, we use a mid $p$-value approach, selecting $\tau = 1/2$ for our work.


To provide further intuition for the the use of conformal win probability, consider a women's basketball game between home team Baylor and away team Oregon State, two highly ranked teams during the 2019-2020 season (see Section \ref{sec:results} for more results related to the top women's teams). We wish to estimate probabilities associated with margins of victory for this particular game. For a specific margin of victory, e.g., a margin of victory of five, $\pi_w(5,\tau)$ is a probability estimate of the event $y_{n+1} \le 5$, which represents a margin of victory of less than or equal to five. Additionally, an estimate for the probability that Baylor wins, i.e., the margin of victory is greater than zero, is $1-\pi_w(0,\tau)$. 

\begin{figure}
    \centering
    \resizebox{.9\textwidth}{!}{
    \includegraphics[trim = 50 0 50 0, clip]{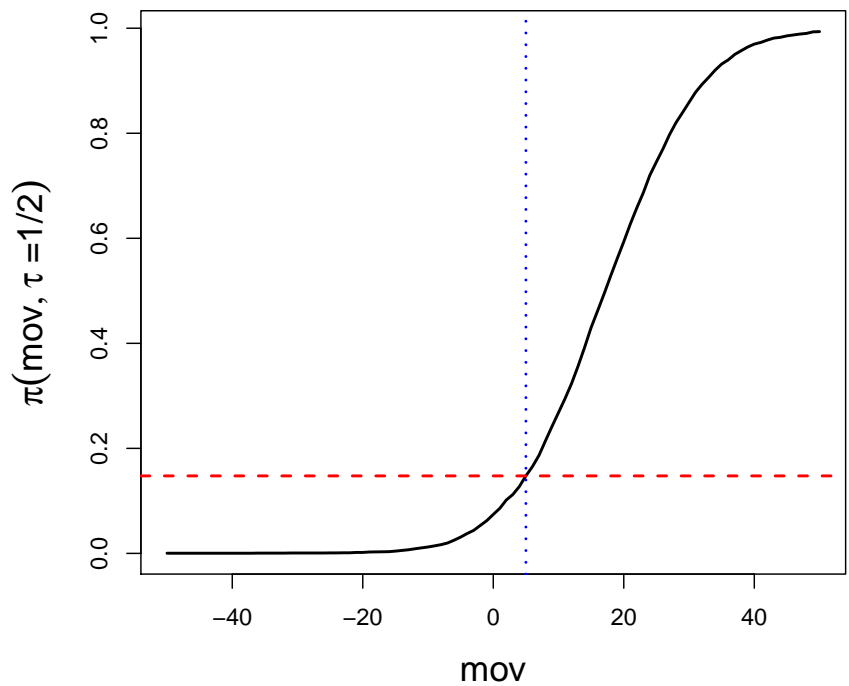}
    }
    \caption{Margin of victory conformal predictive distribution for Baylor vs. Oregon State with $\tau = 1/2$ using regular season data from 2019-2020 NCAA women's basketball season. The blue dotted line identifies a margin of victory for Baylor of 5, i.e., Baylor beating Oregon State by five points, with $\pi(5,1/2) = 0.148$ identified by the red dashed line.}
    \label{fig:cpd-baylor}
\end{figure}

Figure \ref{fig:cpd-baylor} shows the conformal predictive distribution for margin of victory in the case of Baylor vs. Oregon State for the 2019-2020 season. Note that the distribution in Figure \ref{fig:cpd-baylor} has jumps that are too small to be visible. Thus, the distribution is nearly continuous. It is straightforward to reassign probability so that the support of the conformal predictive distribution lies entirely on non-zero integers to match the margin of victory distribution. However, our reassignment does not effect our win probability estimate, so we omit the details here.

With the additional assumptions of mean-zero, independent, normally distributed error terms under \eqref{eqn:lin-model-sports}, the probability construction shown in \eqref{eqn:pivot} becomes

\begin{equation}
\label{eqn:t-wp-sports}
    1 - F_{t,n_w-p}\Bigg(\frac{-\hat{y}_{uvw}}{\hat{\sigma}\sqrt{1+ x'_{uvw}(X_{w-1}'X_{w-1})^{-1}x_{uvw}}}\Bigg),
\end{equation}

\noindent
where $X_w$ is the matrix of covariates up to and including period $w$. 

For logistic regression, we could instead assume

\begin{equation}
\label{eqn:bradley-param2}
\log \bigg( \frac{p_{uvw}}{1-p_{uvw}} \bigg) = x_{uvw}'\beta,
\end{equation}

\noindent
where $p_{uvw}$ is the probability that $y_{uvw}$ is greater than to zero. Then, $p_{uvw}$ is the probability that home team $u$ wins against away team $v$ in period $w$. Similar approaches to \eqref{eqn:bradley-param2} are seen in \cite{bradley1952rank} and \cite{lopez2015building}. The interpretation for $\theta_u - \theta_v$ under model \eqref{eqn:bradley-param2} is no longer the strength difference between teams $u$ and $v$ in terms of MOV, but rather the $\log$-odds of a home team victory when home team $u$ plays away team $v$ at a neutral site. As in linear regression, the rank order of the estimates of the $\theta$ parameters obtained by logistic regression provides a ranking of the teams.

\section{Application to March Madness}
\label{sec:results}

The following section relays the results of the application of conformal win probabilities to the 2019-2020 NCAA Division 1 basketball season. We include estimates of team strengths, probabilities of making the March Madness field, tournament win probabilities, and a comparison of the win probability methods outlined in Section \ref{sec:wp-methods}.

\subsection{Overall Team Strengths for 2019-2020 Season}
\label{sec:ranks}

The regular season ranks and estimated team strengths for the top ten women's and men's teams are shown in Table \ref{tab:end-of-season-ranks-top-women} and Table \ref{tab:end-of-season-ranks-top-men}, respectively. We provide additional 2019-2020 rankings from different sources for comparison, including Associated Press (AP), NCAA Evaluation Tool (NET), KenPom (KP), Ratings Percentage Index (RPI), and College Sports Madness (CSM).

\begin{table}[h]
\centering
\caption{Top 10 NCAA women's teams for 2019-2020 season}
\begin{tabular}{c|c|c|c|c|c}
\hline
        Team & Estimated Strength & Rank & AP & RPI & CSM \\
        \hline
        Baylor & 40.68 & 1 & 3 & 4 & 4 \\
        South Carolina & 40.30 & 2 & 1 & 1 & 1 \\
        Oregon & 39.32 & 3 & 2 & 2 & 2 \\
        Maryland & 37.90 & 4 & 4 & 3 & 6 \\
        Connecticut & 36.17 & 5 & 5 & 4 & 3 \\
        Mississippi St. & 29.07 & 6 & 9 & 10 & 12 \\
        Indiana & 27.91 & 7 & 20 & 14 & 19 \\
        Stanford & 27.82 & 8 & 7 & 6 & 7 \\
        Louisville & 26.36 & 9 & 6 & 7 & 6 \\
        Oregon State & 25.80 & 10 & 14 & 20 & 17 \\
        \hline
\end{tabular}
\label{tab:end-of-season-ranks-top-women}
\end{table}

\begin{table}[h]
\centering
\caption{Top 10 NCAA men's teams for 2019-2020 season}
\begin{tabular}{c|c|c|c|c|c}
\hline
        Team & Estimated Strength & Rank & AP & NET & KP \\
        \hline
        Kansas & 25.26 & 1 & 1 & 2 & 1 \\
        Gonzaga & 22.79 & 2 & 2 & 1 & 2 \\
        Duke & 22.31 & 3 & 11 & 6 & 5 \\
        Michigan State & 20.54 & 4 & 9 & 7 & 7 \\
        Baylor & 20.44 & 5 & 5 & 5 & 3 \\
        Arizona & 19.39 & 6 & - & 14 & 19 \\
        San Diego State & 18.65 & 7 & 6 & 4 & 6 \\
        West Virginia & 18.43 & 8 & 24 & 17 & 10 \\
        Ohio State & 18.22 & 9 & 19 & 16 & 8 \\
        Dayton & 18.07 & 10 & 3 & 3 & 4 \\
        \hline
\end{tabular}
\label{tab:end-of-season-ranks-top-men}
\end{table}

\noindent
The large difference between strengths for the top men's and women's team is due to the difference in team parity between the two leagues, i.e., the gap in strength between the stronger and weaker women's teams is much larger than the gap between the stronger and weaker men's teams.

\subsection{Probabilities of Making March Madness Field for 2019-2020 Season}
\label{sec:adjustment}

The cancellation of the 2020 NCAA basketball post-season prevented the completion of a majority of conference tournaments, as well as the release of final March Madness brackets to the public. At the time of cancellation, there were 20 men's and 18 women's automatic bids still undecided. 

Knowing the results of the (partially) completed conference tournaments allows for estimation of the probabilities of making the March Madness field as outlined in Section \ref{sec:making}. We use regular season data as well as conference tournament progress to update every team's chances of making the tournament at the time of cancellation. Table \ref{tab:women-winners} shows the tournament winners of completed conference tournaments for NCAA women's basketball. These teams have probability 1 of making the March Madness field.

\begin{table}[h]
\caption{Conference champions for 2019-2020 women's basketball season}
\label{tab:women-winners}
\centering
\begin{tabular}{c|c}
\hline
Conference & Winner \\
\hline
Atlantic-10 & Dayton \\
ACC & North Carolina St. \\
American & Connecticut \\
Big East & DePaul \\
Big Ten & Maryland \\
Horizon & IUPUI \\
Ivy League & Princeton$^*$ \\
Mountain West & Boise St. \\
Ohio Valley & Southeast Missouri St. \\
Pac-12 & Oregon \\
SEC & South Carolina \\
Southern & Samford \\
Summit & South Dakota \\
WCC & Portland \\
\hline
\end{tabular}
\end{table}

\noindent
While the Ivy League conference tournament was cancelled, Princeton was awarded an automatic bid to the 2019-2020 March Madness tournament based on their regular season performance. 

With the additional information provided by the outcomes of the completed conference tournaments, there are five different situations for teams as it relates to making the March Madness tournament:

{\singlespacing
\begin{enumerate}
    \item A team has already made the tournament.
    \item A team must win their conference tournament or relies on a small number of teams ranked below them winning their respective conference tournament to make the tournament.
    \item A team has already been eliminated from their conference tournament and relies on a small number of teams ranked below them winning their respective conference tournament to make the tournament.
    \item A team must win their conference tournament to make the tournament.
    \item A team cannot make the tournament.
\end{enumerate}
}

\noindent
Table \ref{tab:women-situations} shows the situations for women's teams ranked from thirty-three to sixty-four.

\begin{table}[h]
    \centering
    \caption{Situations for women's bubble teams}
    \resizebox{\textwidth}{!}{
    \begin{tabular}{c|c}
    \hline
         Situation & Teams \\
         \hline
         1 & Iowa St., Texas, Drake, James Madison, Missouri St., Alabama, TCU, Arizona St., Oklahoma St. \\
         2 & Kansas St. \\
         3 & Marquette, LSU, North Carolina \\
         4 & West Virginia, Oklahoma \\
         5 & all other bubble teams \\
         \hline
    \end{tabular}
    }
    
    \label{tab:women-situations}
\end{table}

\noindent
When using the rankings constructed with regular season data and model \eqref{eqn:lin-model-sports}, the Big 12 conference tournament was the only undecided tournament involving bubble teams, resulting in Kansas State\@ being the sole team in Situation 2 and West Virginia and Oklahoma as the only two teams in Situation 4. Table  \ref{tab:make-tourn-probs} shows the March Madness tournament field probabilities for teams in Situations 2, 3 and 4, constructed with \eqref{eqn:cf-wp} and conformal win probability. Probabilities of making the tournament for the men's teams in Situations 2, 3, and 4 are shown in Supplementary Materials. While not listed in Table \ref{tab:make-tourn-probs}, there is a large number of teams ranked below sixty-four that also fall into Situation 4. 

\begin{table}[h]
\caption{Probabilities of making NCAA tournament field for women's bubble teams for 2019-2020 season.}
\label{tab:make-tourn-probs}
\centering
\begin{tabular}{c|c|c|c}
\hline
Team & Situation & Overall Rank & Probability \\
\hline
Marquette  & 3 & 41 & 0.999 \\
LSU & 3 & 42 & 0.990 \\
North Carolina & 3 & 43 & 0.874 \\
Kansas St. & 2 & 44 & 0.471 \\
West Virginia & 4 & 50 & 0.005 \\
Oklahoma & 4 & 62 & 0.005 \\
\hline
\end{tabular}
\end{table}

\subsection{March Madness Win Probabilities}

Even with the results of the completed conference tournaments, the number of potential tournament brackets remains extremely large. Thus, we forgo the enumeration of all potential brackets and instead focus on three exemplar brackets and three expert brackets to generate March Madness win probabilities. The first two brackets represent two extremes. Bracket 1 maximizes tournament parity, selecting the strongest remaining team from each conference tournament bracket, while Bracket 2 selects the weakest remaining team. Bracket 3 is constructed randomly, selecting teams based on their conference tournament win probabilities. We compare these brackets, and the March Madness win probabilities for the top teams included in these brackets, to those generated by subject matter experts. 

\begin{figure}
    \centering
    \resizebox{\textwidth}{!}{
    \begin{tabular}{cc}
        {\LARGE\textbf{A}} & {\LARGE\textbf{B}} \\ 
        \includegraphics[trim = 0 50 335 100, clip]{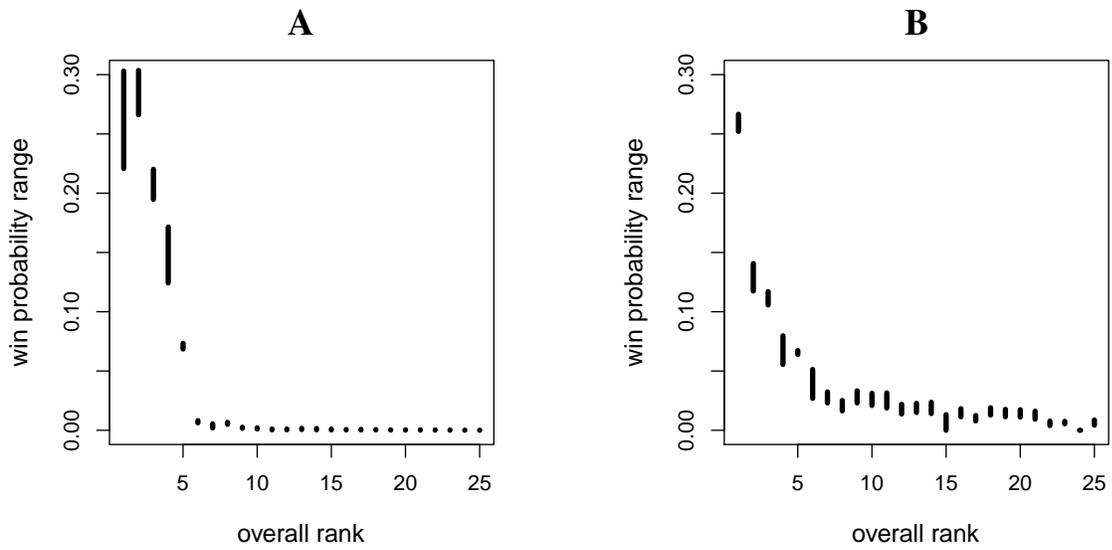} & \includegraphics[trim = 335 50 0 100, clip]{Images/bracket_wp_range_2.eps}  \\
    \end{tabular}
    }
    \caption{Range of exemplar bracket win probabilities for top 25 women's (A) and men's (B) teams during 2019-2020 season.}
    \label{fig:bracket-wp-range}
\end{figure}

\begin{figure}
    \centering
    \resizebox{.9\textwidth}{!}{
    \includegraphics[trim = 100 0 100 0, clip]{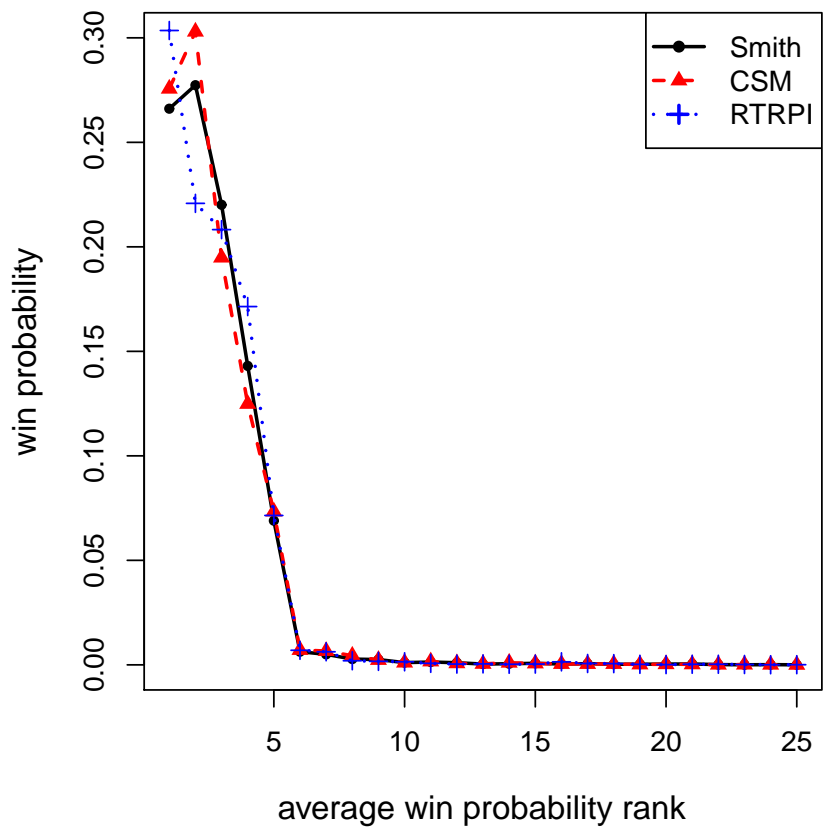}
    }
    \caption{Expert bracket win probabilities for top 25 women's teams.}
    \label{fig:bracket-expert-wp-w}
\end{figure}

\begin{figure}
    \centering
    \resizebox{.9\textwidth}{!}{
    \includegraphics[trim = 100 0 100 0, clip]{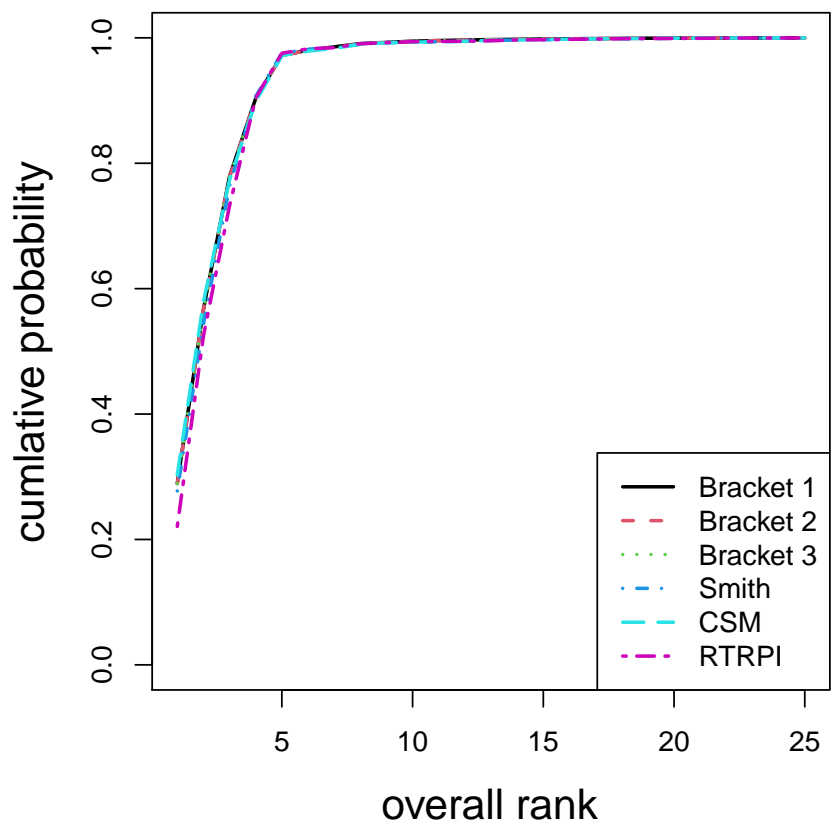}
    }
    \caption{Cumulative tournament win probabilities for top 25 women's teams.}
    \label{fig:women-cdf}
\end{figure}

For the women, we include brackets from basketball expert Michelle Smith \citep{smithbracket2020}, \cite{csmbracket2020} and \cite{rtrpibracket2020}. Table \ref{tab:women-bracket-results} shows the different bracket win probabilities for the top ten women's teams, ranked using the ranking method outlined in Section \ref{sec:sports-app}. Exemplar bracket results for the men's 2019-2020 season are shown in Supplementary Materials, with brackets generated by NCAA basketball experts Andy Katz \citep{katzbracket2020}, Joe Lunardi \citep{lunardiracket2020} and Jerry Palm \citep{palmbracket2020}. Figure \ref{fig:bracket-wp-range} shows the ranges of win probabilities across all exemplar brackets for the top 25 teams. Figure \ref{fig:bracket-expert-wp-w} shows a comparison of win probabilities across the expert generated brackets. Figure \ref{fig:women-cdf} compares cumulative NCAA tournament win probabilities across brackets for the top 25 women's teams. The cumulative NCAA tournament win probabilities for the top 25 men's teams are included in Supplementary Materials. 

\begin{table}[h]
    \centering
    \caption{March Madness win probabilities given exemplar brackets for top ranked women's teams.}
    \resizebox{.8\textwidth}{!}{
    \begin{tabular}{c|c|c|c|c|c|c}
    \hline
    Team & Bracket 1 & Bracket 2 & Bracket 3 & Smith & CSM & RTRPI \\
    \hline
    Baylor & 0.289 & 0.289 & 0.289 & 0.277  & 0.303 & 0.221 \\
    South Carolina & 0.278 & 0.277 & 0.278 & 0.267 & 0.276 & 0.304 \\
    Oregon & 0.212 & 0.212 & 0.212 & 0.220 & 0.195 & 0.208 \\
    Maryland & 0.124 & 0.125 & 0.124 & 0.143 & 0.125 & 0.171 \\
    Connecticut & 0.069 & 0.069 & 0.069 & 0.069 & 0.073 & 0.071 \\
    Mississippi St. & 0.008 & 0.008 & 0.008 & 0.006 & 0.007 & 0.007 \\
    Indiana & 0.005 & 0.005 & 0.005 & 0.003 & 0.004 & 0.002 \\
    Stanford & 0.005 & 0.005 & 0.005 & 0.005 & 0.007 & 0.006 \\
    Louisville & 0.002 & 0.002 & 0.002 & 0.003 & 0.002 & 0.002 \\
    Oregon St. & 0.002 & 0.002 & 0.002 & 0.001 & 0.001 & 0.001 \\
    \hline
    \end{tabular}
    }
    \label{tab:women-bracket-results}
\end{table}

\begin{figure}
    \centering
    \resizebox{\textwidth}{!}{
    \includegraphics{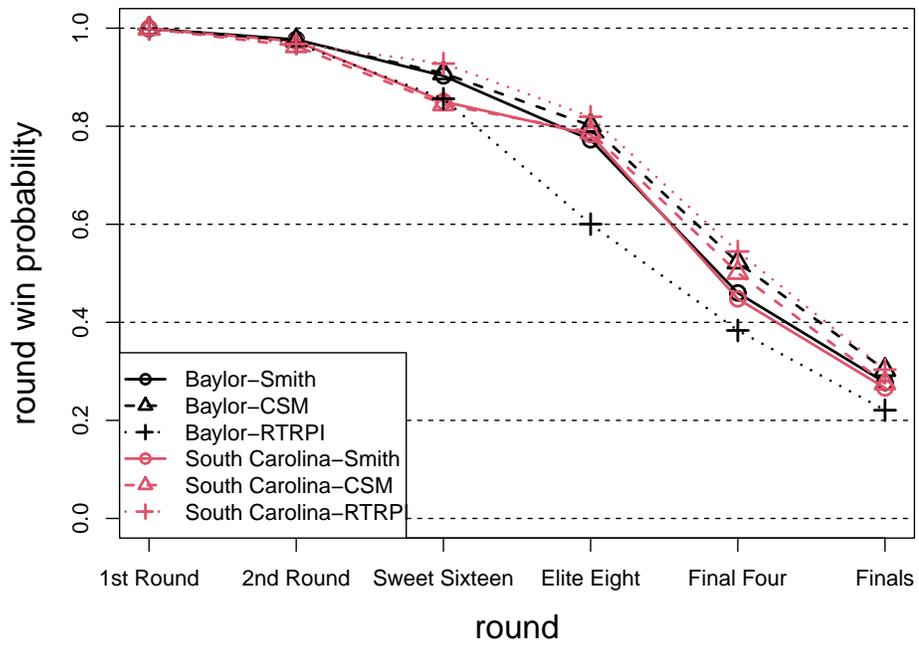}
    }
    \caption{Round-by-round win probabilities for Baylor and South Carolina. The values shown indicate the probabilities of a team moving on from a particular round.}
    \label{fig:rbr-wp-w}
\end{figure}

In general, tournament probabilities do not change drastically across brackets. However, we do see larger probability ranges associated with the top women's teams. Specifically, the tournament win probability for Baylor, the highest ranked team with respect to our ranking, drops to 0.221 with the RTRPI expert bracket, as opposed to 0.288 and 0.303 for the Smith and CSM brackets, respectively. Additionally, the overall tournament win probability for South Carolina increases to 0.304 with the RTRPI bracket. Figure \ref{fig:rbr-wp-w} shows round-by-round win probabilities for Baylor and South Carolina for each of the expert brackets.  

We see that Baylor's RTRPI round-by-round win probability becomes lower than South Carlolina's after the second round, dropping to 0.856, compared to South Carlolina's 0.927. The largest decrease occurs during the Elite Eight, where Baylor's probability of moving on from the Elite Eight (under the RTRPI bracket) is 0.600, compared to South Carolina's 0.819. This is due to Connecticut's placement in the same region as Baylor, with each team seeded as the 1-seed and 2-seed, respectively. In the other expert brackets, Connecticut was placed in the same region as Maryland. The other brackets keep the round-by-round win probabilities for these two teams relatively stable.

\subsection{Win Probability Calibration}
\label{sec:dfpi-effectiveness}

In order to assess the win probability estimates generated using the methods outlined in Section \ref{sec:wp-methods}, we compare estimates for previous NCAA basketball seasons, including the shortened 2019-2020 season. We use the regular season games to estimate the team strengths and then construct win probabilities for each game of post-season play. 

\begin{figure}
    \centering
    \resizebox{\textwidth}{!}{
    \includegraphics[trim = 0 0 0 20, clip]{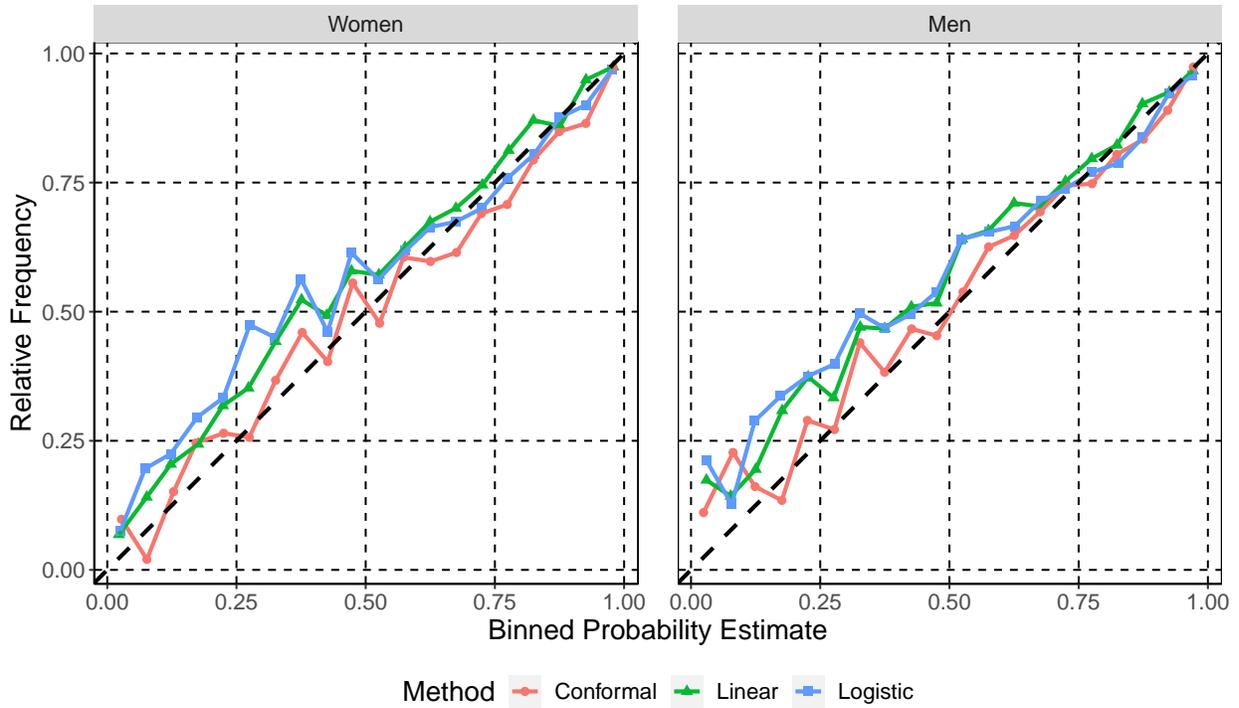}
    }
    \caption{Empirical calibration comparison for NCAA women's and men's basketball for 2014-2015 to 2020-2021 post-seasons for methods outlined in Section \ref{sec:wp-methods}.}
    \label{fig:calibration-all-byhome}
\end{figure}

Ideally, the estimated probability for an event occurring should be \textit{calibrated}. A perfectly calibrated model is one such that

\begin{equation}
\label{eqn:calibration}
\textrm{E}_{\hat{p}} \Big[\bigg|P \Big( \hat{z} = z|\hat{p} = p \Big) - p \bigg| \Big] = 0,
\end{equation}

\noindent
where $z$ is an observed outcome, $\hat{z}$ is the predicted outcome, $\hat{p}$ is a probability estimate for the predicted outcome, and $p$ is the true outcome probability \citep{guo2017calibration}. In the NCAA basketball case \eqref{eqn:calibration} implies that if we inspect, say, each game with an estimated probability of 40\% for home team victory, we should expect a home team victory in 40\% of the observed responses. We can assess calibration in practice by grouping similarly valued probability estimates into a single bin and then calculating the relative frequency of home team victories for observations within each bin. For visual comparison of calibration, Figure \ref{fig:calibration-all-byhome} shows a reliability plot for the win probability estimates generated using the methods outlined in Section \ref{sec:wp-methods} with bin intervals of width 0.025. From Figure \ref{fig:calibration-all-byhome} we can see that while the methods are comparable for higher win probability estimates, the conformal win probability approach is much better calibrated for lower win probability estimates. A majority of observed relative frequencies for conformal win probabilities fall closer to the dotted line, signifying better calibration than the other two methods.

\begin{figure}
    \centering
    
    \resizebox{\textwidth}{!}{
    \includegraphics[trim = 0 0 0 0, clip]{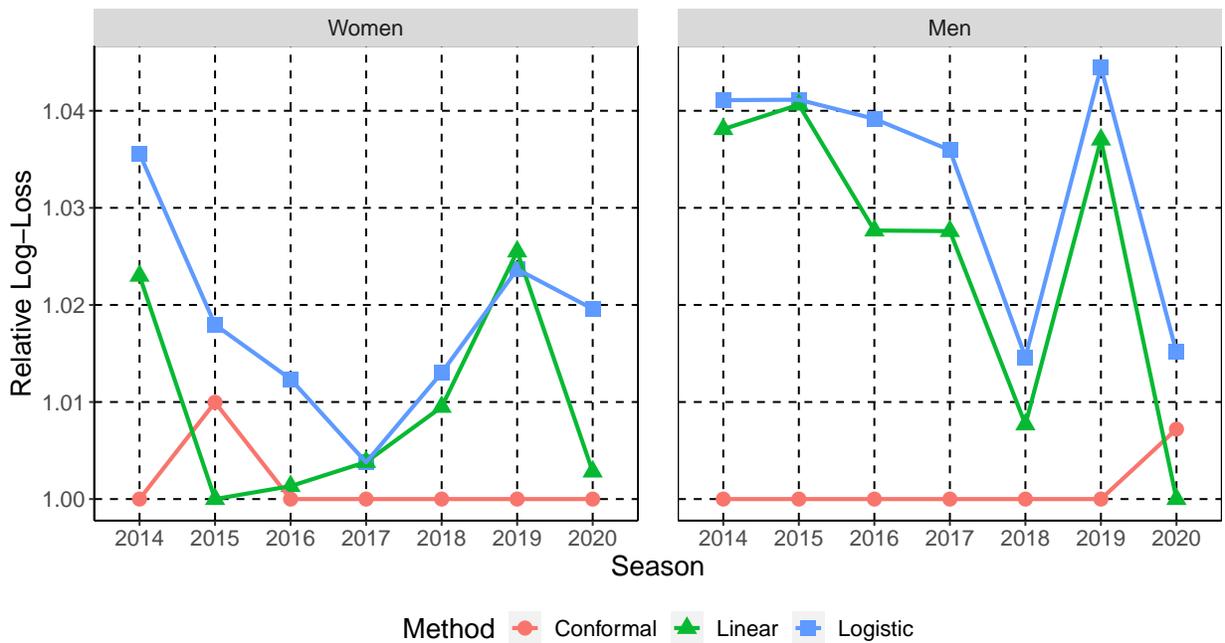}
    }
    \caption{Relative $\log$-loss comparison for NCAA women's and men's basketball for win probability estimates associated with 2014-2015 to 2020-2021 post-seasons for methods outlined in Section \ref{sec:wp-methods}.}
    \label{fig:log-loss}
\end{figure}

To provide a numerical interpretation of calibration, we compare the three probability estimation approaches mentioned in Section \ref{sec:wp-methods} using $\log$-loss

\begin{equation}
    \log L(\hat{p}, z) = z\log(\hat{p}) + (1-z)\log(1-\hat{p}),
\end{equation}

\noindent
which generates loss for each individual win probability estimate rather than a group of binned estimates. $\log$-loss has been shown to have strong empirical and theoretical properties as a loss function \citep{painsky2018universality, vovk2015fundamental}. Figure \ref{fig:log-loss} shows the \textit{relative} $\log$-loss, i.e., the ratio of the $\log$-loss for one method to the minimum $\log$-loss across all methods, broken up by season and league.


We see that for all year-league combinations except for the women's 2015-2016 season and men's 2020-2021 season, conformal win probabilities performed better than the other two methods. Additionally, even when conformal win probabilities are not the best performing approach, they still result in a $\log$-loss within one percent of the best performing approach. Table \ref{tab:log-loss-all} shows the results for the entire collection of probability estimates for each league.

\begin{table}[h]
\centering
\caption{Relative $\log$-loss for NCAA men's and women's basketball win probability estimates by league.}
\label{tab:log-loss-all}
\begin{tabular}{c|c|c|c}
\hline
 & \multicolumn{3}{c}{Method} \\
 \hline
League & Conformal & Linear & Logistic \\
\hline
Women & 1.00 & 1.01 & 1.02 \\
Men & 1.00 & 1.02 & 1.03 \\
\hline
\end{tabular}
\end{table}

\section{Conclusion}
\label{sec:conclusion}

The cancellation of March Madness in 2020 resulted in disappointment for many across the country, fans and athletes alike. We explored win probabilities as they relate to the NCAA tournament, delivering a closed-form calculation for probabilities of making the tournament, given a set of team strengths estimated from game outcomes. We introduced conformal win probabilities and compared to win probabilities derived from logistic regression and linear regression assuming normally distributed, independent, mean-zero errors. Conformal win probabilities were superior to those obtained from the other methods.

For the application in this paper, we limited our discussion to model \eqref{eqn:lin-model-sports1}. Each of the win probability methods described in Section \ref{sec:wp-methods} can be applied to more complex models, so future work could focus on comparing these methods in a more complex setting. One example of a model we could assume is

\begin{equation}
y_{uvw} = \mu + \theta_{uw} - \theta_{vw} + \epsilon_{uvw},
\label{eqn:lin-model-sports-fused} 
\end{equation}

\noindent
where $\theta_{uw}$ is the strength of team $u$ during week $w$. \eqref{eqn:lin-model-sports-fused} is rank deficient, so we could consider a fused lasso approach \citep{tibshirani2005sparsity}, where the objective function is penalized by $\lambda \sum_u \sum_{w = 1}^{W-1} |\theta_{uw} - \theta_{uw+1}|$ to encourage the difference in parameter values from one period to the next to be small for each team. This approach allows for relative team strengths to change across a season, rather than estimating one average strength for each team over the course of the entire season. Additionally, we could incorporate team ``match-up" statistics, e.g., the difference between the teams' offensive or defensive efficiencies, rather than solely estimating a win probability based on the teams playing.

The focus on event probabilities can also be extended to a betting scenario. In this paper, the event probability of interest was a win (or loss) for a specific team. This event corresponds to a ``moneyline" bet in sports betting, i.e., betting on a specific team to win a game. Another type of bet is the ``spread" bet, which accounts for differences in the strengths of two teams, either through the adjustment of a point spread or the odds associated with a particular team. The spread is chosen by bookmakers so that the total amount of money bet on the spread of the favorite is near that bet against favorite (as opposed to being representative of, say, the expected margin of victory).  For example, suppose we have an upcoming contest between two teams, a favorite and an underdog, with a spread of negative three. A bettor taking the spread on the favorite would win the bet if the favorite wins by more than three points, while a bettor taking the spread against the favorite would win the bet if the underdog wins or loses by less than three points. In order to determine whether to bet on the favorite or the underdog in a spread bet, we can utilize conformal win probabilities. Specifically, calculating $\pi(-s,1/2)$, where $s$ is the spread for a game of interest, generates an estimate of the probability that the margin of victory (favorite score - underdog score) will be less than or equal to $-s$.

One other major simplification we utilize in this paper is that estimated team strength does not change following the regular season. Thus, we eliminate the potential for teams to receive a higher (or lower) overall rank based on their conference tournament performance. While this simplifies the analysis, allowing for teams to move up or down in rank might more closely match the March Madness selection committee's actual process. 

\bibliographystyle{apalike}
\bibliography{main}

\clearpage

\clearpage
\beginsupplement
\begin{figure}
    \centering
    \resizebox{\textwidth}{!}{
    \includegraphics[trim = 100 0 100 0, clip]{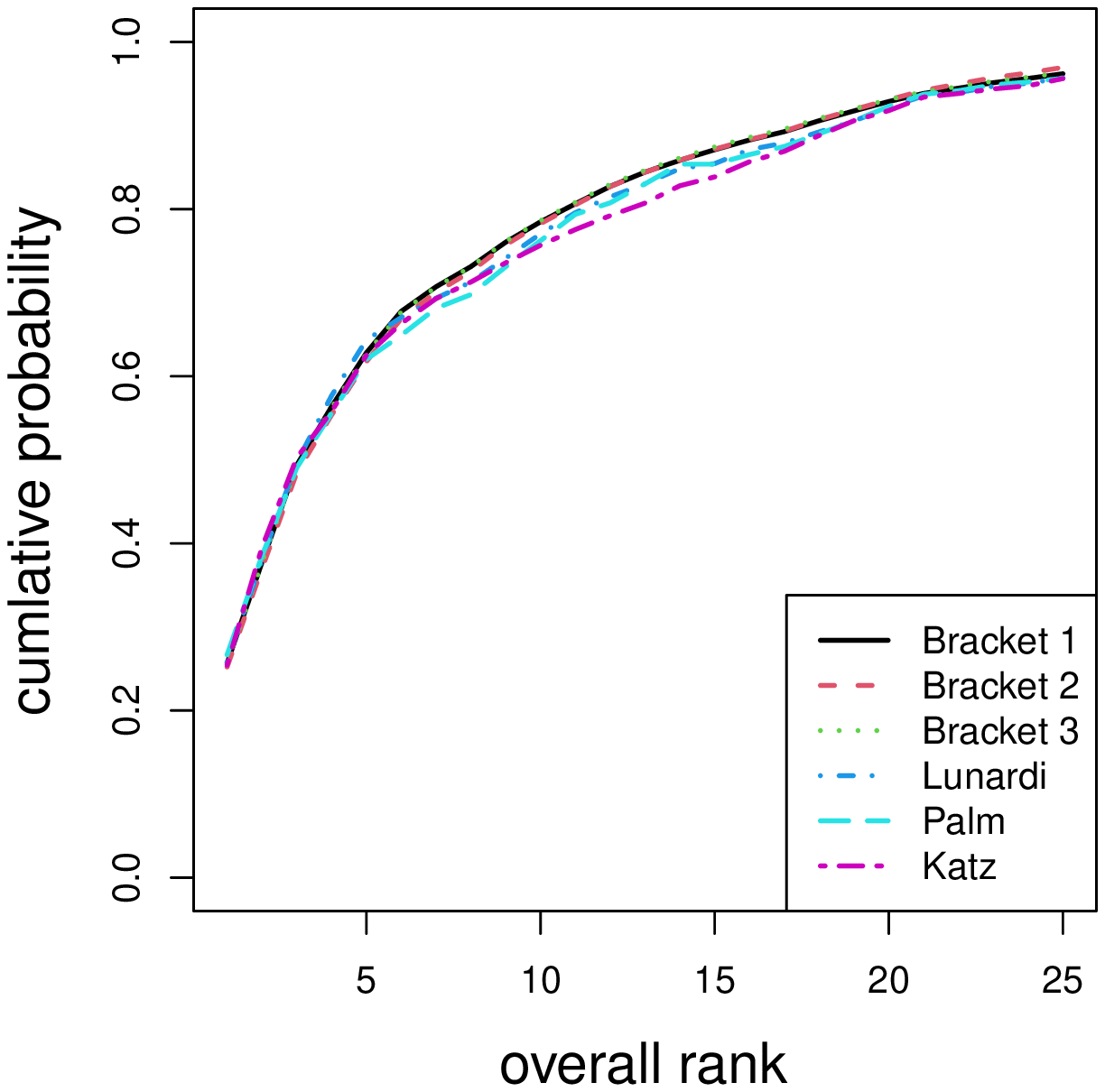}
    }
    \caption{Cumulative tournament win probabilities for top 25 men's teams.}
    \label{fig:men-cdf}
\end{figure}

\section*{Supplementary Materials}
\beginsupplement

\subsection{Closed-Form Probability Calculations for March Madness Rank}
\label{sec:mm-ranks}

A team's probability of winning March Madness is not only dependent on making the tournament, but also on where they are ranked within the tournament. Each team's rank in the tournament, as well as the selection of the thirty-two at-large bids, are decided by a selection committee. The sixty-four team field is partitioned into four regions, each with sixteen teams. Within a region, each team is assigned a ``seed" according to their rank. Stronger teams receive lower seeds, while weaker teams receive higher seeds. Ideally, the strongest team in each region receives a 1-seed, the second strongest a 2-seed, and so on. There are additional constraints related to match-ups between teams in the same conferences as well as location considerations \citep{ncaa2021}. 

To keep each region equally competitive, each of the four teams at each seed are placed into different regions using the S-curve method. The S-curve method places the teams in regions one through four based on their overall rank according to the order shown in Table \ref{tab:s-curve}. While a simplification of the actual selection process, we use the S-curve method, combined with estimated teams strengths constructed using the method outlined in Section \ref{sec:sports-app}, as the sole tools for constructing a bracket given a tournament field.

\begin{table}[h]
\caption{Region placement using S-curve method}
\label{tab:s-curve}
\centering
\begin{tabular}{c|c|c|c|c}
\hline
 & \multicolumn{4}{c}{Region} \\
 \hline
Seed & 1 & 2 & 3 & 4 \\
\hline
1 & 1 & 2 & 3 & 4 \\
2 & 8 & 7 & 6 & 5 \\
3 & 9 & 10 & 11 & 12 \\
4 & 16 & 15 & 14 & 13 \\
5 & 17 & 18 & 19 & 20 \\
6 & 24 & 23 & 22 & 21 \\
7 & 25 & 26 & 27 & 28 \\
8 & 32 & 31 & 30 & 29 \\
9 & 33 & 34 & 35 & 36 \\
10 & 40 & 39 & 38 & 37 \\
11 & 41 & 42 & 43 & 44 \\
12 & 48 & 47 & 46 & 45 \\
13 & 49 & 50 & 51 & 52 \\
14 & 56 & 55 & 54 & 53 \\
15 & 57 & 58 & 59 & 60 \\
16 & 64 & 63 & 62 & 61 \\
\hline
\end{tabular}
\end{table}

As in Section \ref{sec:making}, let team $u$ be the team whose overall rank is $u$. We define $R_u$ as a random variable describing the March Madness tournament rank for team $u$. Then, the probability of team $u$ being ranked $r$ in the tournament and making the tournament field, can be decomposed into

\begin{equation}
\label{eqn:wp-total-prob2}
\mathbb{P}(R_u= r, F_u = 1) = \mathbb{P}(R_u = r|F_u = 1)\mathbb{P}(F_u = 1).
\end{equation}

\noindent
The conditional portion of \eqref{eqn:wp-total-prob2} can be further decomposed into

\begin{equation}
\label{eqn:rank-probs}
\mathbb{P}(R_u = r|F_u = 1) = \left\{
        \begin{array}{c@{}c@{}}
            0 & \quad r > u \\
            \mathbb{P}(\{L_u \le t_r\} \triangle \{C_u = 1\}) & \quad r = u \\
            \mathbb{P}(\{L_u = t_{r-1}\} \cap \{C_u = 1\}) & \quad r < u
        \end{array}\; ,
    \right. 
\end{equation}

\noindent
where $A \triangle B$ is the symmetric difference between events $A$ and $B$. The symmetric difference between the events $\{L_u \le t_u\}$ and $\{C_u = 1\}$ captures the scenario where team $u$ either wins their conference tournament or makes the March Madness field as an at-large bid, but not both. With our simplifications, a team cannot be ranked lower in the tournament than their overall rank, so a March Madness tournament rank $r > u$ occurs with probability zero. To receive a rank higher than their overall rank, a team must win their conference tournament, and (any number of) teams ranked ahead of them must be eliminated from field contention.

The only way into the tournament for low-ranked teams is to win their conference tournament. Tournament ranks for these teams are dependent on how many other teams ranked higher than them win their respective conference tournament. The conditional probability of tournament rank $r$ for low-ranked team $u$ is

\begin{equation}
\label{eqn:wp-total-prob3}
\mathbb{P}(R_u = r|F_u = 1) =  \mathbb{P}\Bigg(\sum_{k = 1}^K C_{\mathcal{H}^u_k}= r - 32 \Bigg),
\end{equation}

\noindent
where $\{C_{\mathcal{H}^u_{k(u)}} = 1\}$ occurs with probability one because we condition on team $u$ having already won their conference tournament. 

\subsection{Additional Tables and Figures}

\begin{table}[h]
    \centering
    \caption{Situations for men's bubble teams}
    \resizebox{\textwidth}{!}{
    \begin{tabular}{c|c}
         Situation & Teams \\
         \hline
         1 & Utah St., Florida, Auburn \\
         2 & Indiana, LSU, Arkansas, Oklahoma, Wichita St., Cincinnati \\
         3 & Stanford \\
         4 & Alabama, Providence, Syracuse, Mississippi St., Memphis, NC St., Arizona St., Rhode Island, \\
         & Virginia, USC, Oklahoma St., Tennessee, Notre Dame, Richmond, Yale, Clemson, Connecticut \\
         5 & all other teams \\
         \hline
    \end{tabular}
    }
    
    \label{tab:mens-situations}
\end{table}

\begin{table}[h]
\caption{Probabilities of making NCAA tournament field for men's bubble teams for 2019-2020 season.}
\label{tab:make-tourn-probs-m}
\centering
\begin{tabular}{c|c|c|c}
\hline
Team & Situation & Overall Rank & Probability \\
\hline
Oklahoma & 2 & 40 & 0.999 \\
Wichita St. & 2 & 41 & 0.999 \\
Cincinnati & 2 & 42 & 0.997 \\
Xavier & 3 & 43 & 0.970 \\
St. Mary's (CA) & 3 & 44 & 0.846\\
Alabama & 2 & 45 & 0.575 \\
Providence & 2 & 46 & 0.231 \\
Syracuse & 2 & 47 & 0.040 \\
Mississippi St. & 4 & 48 & 0.082 \\
Memphis & 4 & 49 & 0.114 \\
NC St. & 4 & 50 & 0.021 \\
Arizona St. & 4 & 51 & 0.141 \\
Rhode Island & 4 & 52 & 0.125 \\
Virginia & 4 & 53 & 0.041 \\
USC & 4 & 56 & 0.041 \\
Oklahoma St. & 4 & 57 & 0.007 \\
Tennessee & 4 & 58 & 0.026 \\
Notre Dame & 4 & 60 & 0.033 \\
Richmond & 4 & 61 & 0.104 \\
Yale & 4 & 62 & 0.520 \\
Clemson & 4 & 63 & 0.017 \\
Connecticut & 4 & 64 & 0.064 \\
Texas & 4 & 65 & 0.008 \\
VCU & 4 & 66 & 0.053 \\
Davidson & 4 & 67 & 0.060 \\
South Carolina & 4 & 68 & 0.020 \\
\hline
\end{tabular}
\end{table}

\begin{table}[h]
\caption{Bracket 1 for women's tournament and respective seeding}
\label{tab:bracket-w-max}
\centering
{\footnotesize
\begin{tabular}{c|c|c|c|c}
\hline
Seed & Region 1 & Region 2 & Region 3 & Region 4 \\
\hline
1 & Baylor & Maryland & South Carolina & Oregon \\
2 & Stanford & Connecticut & Indiana & Mississippi St. \\
3 & Louisville & Arkansas & Oregon St. & South Dakota \\
4 & Arizona & Northwestern & NC St. & Princeton \\
5 & UCLA & Texas A\&M & Iowa & Kentucky \\
6 & Gonzaga & Florida St. & Ohio St. & DePaul \\
7 & Rutgers & Virginia Tech & Georgia Tech & Tennessee \\
8 & Iowa St. & Florida Gulf Coast & Duke & Michigan \\
9 & Texas & Missouri St. & Drake & JamesMadison \\
10 & Oklahoma St. & Alabama & Arizona St. & TCU \\
11 & Marquette & Kansas St. & LSU & North Carolina \\
12 & Ohio & IUPUI & Dayton & Montana St. \\
13 & Old Dominion & Troy & Bucknell & Marist \\
14 & SE Missouri St. & Stephen F. Austin & Portland & Boise St. \\
15 & Robert Morris & Stony Brook & UC Davis & Texas Southern \\
16 & Norfolk St. & Samford & Kansas City & Campbell \\
\hline
\end{tabular}
}
\end{table}

\begin{table}[h]
\caption{Bracket 2 for women's tournament and respective seeding}
\label{tab:bracket-w-min}
\centering
{\footnotesize
\begin{tabular}{c|c|c|c|c}
\hline
Seed & Region 1 & Region 2 & Region 3 & Region 4 \\
\hline
1 & Baylor & Maryland & South Carolina & Oregon \\
2 & Stanford & Connecticut & Indiana & Mississippi St. \\
3 & Louisville & Arkansas & Oregon St. & South Dakota \\
4 & Arizona & Northwestern & NC St. & Princeton \\
5 & UCLA & Texas A\&M & Iowa & Kentucky \\
6 & Gonzaga & Florida St. & Ohio St. & DePaul \\
7 & Rutgers & Virginia Tech & Georgia Tech & Tennessee \\
8 & Iowa St. & Florida Gulf Coast & Duke & Michigan \\
9 & Texas & Missouri St. & Drake & James Madison \\
10 & Oklahoma St. & Alabama & Arizona St. & TCU \\
11 & IUPUI & Boise St. & Dayton & Kansas \\
12 & Idaho & Portland & Liberty & SE Missouri St. \\
13 & Toledo & South Alabama & Samford & Boston \\
14 & NC Wilmington & Maine & Alabama A\&M & Marshall \\
15 & Fairfield & Grand Canyon & Indiana St. & Cal Poly SLO \\
16 & USC Upstate & Incarnate Word & MD Eastern Shore & Fairleigh Dickinson \\
\hline
\end{tabular}
}
\end{table}

\begin{table}[h]
\caption{Bracket 3 for women's tournament and respective seeding}
\label{tab:bracket-w-random}
\centering
{\footnotesize
\begin{tabular}{c|c|c|c|c}
\hline
Seed & Region 1 & Region 2 & Region 3 & Region 4 \\
\hline
1 & Baylor & Maryland & South Carolina & Oregon \\
2 & Stanford & Connecticut & Indiana & Mississippi St. \\
3 & Louisville & Arkansas & Oregon St. & South Dakota \\
4 & Arizona & Northwestern & NC St. & Princeton \\
5 & UCLA & Texas A\&M & Iowa & Kentucky \\
6 & Gonzaga & Florida St. & Ohio St. & DePaul \\
7 & Rutgers & Virginia Tech & Georgia Tech & Tennessee \\
8 & Iowa St. & Florida Gulf Coast & Duke & Michigan \\
9 & Texas & Missouri St. & Drake & James Madison \\
10 & Oklahoma St. & Alabama & Arizona St. & TCU \\
11 & Marquette & Kansas St. & LSU & North Carolina \\
12 & Ohio & IUPUI & Dayton & Montana St. \\
13 & Rice & Boise St. & Marist & Coastal Carolina \\
14 & Lehigh & Portland & Robert Morris & SE Missouri St. \\
15 & Abilene Christian & Maine & Hawaii & Samford \\
16 & MD Eastern Shore & Alabama A\&M & Seattle & Radford \\
\hline
\end{tabular}
}
\end{table}

\begin{table}[h]
\caption{Bracket 1 for men's tournament and respective seeding}
\label{tab:bracket-m-max}
\centering
{\footnotesize
\begin{tabular}{c|c|c|c|c}
\hline
Seed & Region 1 & Region 2 & Region 3 & Region 4 \\
\hline
1 & Kansas & Gonzaga & Duke & Michigan St. \\
2 & West Virginia & San Diego St. & Arizona & Baylor \\
3 & Ohio St. & Dayton & Maryland & Michigan \\
4 & Creighton & Texas Tech & Florida St. & Louisville \\
5 & BYU & Oregon & Seton Hall & Villanova \\
6 & Marquette & Iowa & Houston & Penn St. \\
7 & Colorado & Purdue & Kentucky & Wisconsin \\
8 & Butler & Rutgers & Minnesota & Illinois \\
9 & Utah St. & Florida & Auburn & Indiana \\
10 & Oklahoma & Stanford & Arkansas & LSU \\
11 & Wichita St. & Cincinnati & Xavier/Providence &  St. Mary's (CA)/Alabama \\
12 & Liberty & North Texas & E. Tennessee St. & Yale \\
13 & Vermont & Akron & N Colorado & Belmont \\
14 & Bradley & UC Irvine & Stephen F. Austin & Texas St. \\
15 & New Mexico St. & Hofstra & Winthrop & North Dakota St. \\
16 & Prairie View A\&M/Robert Morris & Siena/Norfolk St. & Boston & Northern Kentucky \\
\hline
\end{tabular}
}
\end{table}

\begin{table}[h]
\caption{Bracket 2 for men's tournament and respective seeding}
\label{tab:bracket-m-min}
\centering
{\footnotesize
\begin{tabular}{c|c|c|c|c}
\hline
Seed & Region 1 & Region 2 & Region 3 & Region 4 \\
\hline
1 & Kansas & Gonzaga & Duke & Michigan St. \\
2 & West Virginia & San Diego St. & Arizona & Baylor \\
3 & Ohio St. & Dayton & Maryland & Michigan \\
4 & Creighton & Texas Tech & Florida St. & Louisville \\
5 & BYU & Oregon & Seton Hall & Villanova \\
6 & Marquette & Iowa & Houston & Penn St. \\
7 & Colorado & Purdue & Kentucky & Wisconsin \\
8 & Butler & Rutgers & Minnesota & Illinois \\
9 & Utah St. & Florida & Auburn/Stanford & Indiana \\
10 & Liberty & E Tennessee St. & Clemson & LSU/Arkansas \\
11 & DePaul & Kansas St. & Belmont & Bradley \\
12 & Texas A\&M & North Dakota St. & Winthrop & Hofstra \\
13 & California & Northern Kentucky & South Alabama & Boston \\
14 & Robert Morris & E. Carolina & Florida Atlantic & Princeton \\
15 & Fordham & Miami Ohio & Hartford & UT Rio Grande Valley \\
16 & Niagara/Northwestern St. & Jackson St./S Carolina St. & Long Beach St. & Idaho St. \\
\hline
\end{tabular}
}
\end{table}

\begin{table}[h]
\caption{Bracket 3 for men's tournament and respective seeding}
\label{tab:bracket-m-random}
\centering
{\footnotesize
\begin{tabular}{c|c|c|c|c}
\hline
Seed & Region 1 & Region 2 & Region 3 & Region 4 \\
\hline
1 & Kansas & Gonzaga & Duke & Michigan St. \\
2 & West Virginia & San Diego St. & Arizona & Baylor \\
3 & Ohio St. & Dayton & Maryland & Michigan \\
4 & Creighton & Texas Tech & Florida St. & Louisville \\
5 & BYU & Oregon & Seton Hall & Villanova \\
6 & Marquette & Iowa & Houston & Penn St. \\
7 & Colorado & Purdue & Kentucky & Wisconsin \\
8 & Butler & Rutgers & Minnesota & Illinois \\
9 & Utah St. & Florida & Auburn & Indiana \\
10 & Oklahoma & Stanford & Arkansas & LSU \\
11 & Wichita St. & Cincinnati & Xavier/Providence &  St. Mary's (CA)/Alabama \\
12 & Liberty & North Texas & E. Tennessee St. & Yale \\
13 & Vermont & Belmont & Stephen F. Austin & Bradley \\
14 & Winthrop & Eastern Washington & Hofstra & New Mexico St. \\
15 & North Dakota St. & Northern Kentucky & South Alabama & Boston \\
16 & Norfolk St./CS Fullerton & Robert Morris/Southern & Siena & N. Illinois \\
\hline
\end{tabular}
}
\end{table}

\begin{table}[h]
    \centering
    \caption{March Madness win probabilities given exemplar brackets for top ranked men's teams.}
    \resizebox{.8\textwidth}{!}{
    \begin{tabular}{c|c|c|c|c|c|c}
    \hline
    Team & Bracket 1 & Bracket 2 & Bracket 3 & Lunardi & Palm & Katz \\
    \hline
    Kansas & 0.256 & 0.252 & 0.256 & 0.258 & 0.267 & 0.254 \\
    Gonzaga & 0.121 & 0.117 & 0.121 & 0.122 & 0.118 & 0.141 \\
    Duke & 0.116 & 0.115 & 0.116 & 0.117 & 0.106 & 0.108 \\
    Michigan St. & 0.070 & 0.069 & 0.070 & 0.080 & 0.065 & 0.056 \\
    Baylor & 0.065 & 0.064 & 0.065 & 0.067 & 0.065 & 0.067 \\
    Arizona & 0.050 & 0.051 & 0.050 & 0.027 & 0.029 & 0.038 \\
    San Diego St. & 0.030 & 0.032 & 0.030 & 0.023 & 0.032 & 0.030 \\
    West Virginia & 0.024 & 0.025 & 0.024 & 0.019 & 0.016 & 0.020 \\
    Ohio St. & 0.029 & 0.032 & 0.030 & 0.029 & 0.033 & 0.023 \\
    Dayton & 0.024 & 0.025 & 0.025 & 0.029 & 0.031 & 0.021 \\
    \hline
    \end{tabular}
    }
    \label{tab:men-bracket-results}
\end{table}
\end{document}